\documentclass[preprint,aps,nofootinbib]{revtex4}

\newcommand {\bc}{\begin{center}}
\newcommand {\ec}{\end{center}}

\def\lsim{\mathrel{\rlap{\lower4pt\hbox{$\sim$}}
    \raise1pt\hbox{$<$}}}               
\def\gsim{\mathrel{\rlap{\lower4pt\hbox{$\sim$}}
    \raise1pt\hbox{$>$}}}  
              
\usepackage{graphicx}

\begin{document}

\title{Quantum limited sound attenuation in a dilute atomic
Fermi gas: Shaken, not stirred}

\author{Thomas Sch\"afer}

\affiliation{Department of Physics, 
North Carolina State University, 
Raleigh, NC 27695}

\begin{abstract}
A new experiment involving resonantly interacting atoms 
confined by laser beams sheds light on momentum and energy 
diffusion in quantum fluids.
\end{abstract}

\maketitle

 Ordinary sound is a harmonic oscillation in the density, temperature, and 
velocity of air. Sound intensity decreases because of the spreading of the 
sound wave, but ultimately sound attenuation is due to diffusion of momentum 
and energy from the crest to the trough of the wave. This effect can be 
characterized in terms of the diffusivity $D$ of sound. In air there is 
a very large separation of scales between the shortest scale, the distance
between molecules, an intermediate scale, the mean free path of air molecules
which controls the diffusivity, and the longest scale, the wavelength of
the sound mode. A very different, deeply quantum, version of sound
attenuation was studied by Patel et al.~and is described on page 1222
of  Science 370 (2020) 6521 \cite{Patel:2019udb}. Their result sheds
light on transport properties of strongly correlated quantum fluids
\cite{Schafer:2009dj}, and has direct implications for the stability
of spinning neutron stars \cite{Alford:2010fd}. 

In the experiment of Patel et al.~about two million Lithium atoms are
confined in a cylindrical box created by beams of laser light, see Fig.~1.
The box is about 100 $\mu m$ long, and 60 $\mu m$ in radius. A typical 
standing wave studied in the experiment has a wave length which is
only about ten times larger than the mean distance between atoms. In order 
to observe sharp collective modes in this regime the gas must be very
strongly correlated. This is achieved by making the gas very cold, and
by tuning the interaction between atoms to a resonance. The temperature
of the gas is between 50 and 500 nK, which implies that the de Broglie 
wave length of the atoms is equal to or larger than the mean atomic
distance. Here, the de Broglie wave length is the wave length of
quantum mechanical wave function of the atoms. The interaction between
the atoms is tuned by means of a so-called Feshbach resonance
\cite{Bloch:2008zzb}. On resonance we can think of the interaction
as having zero range, but infinite scattering length. This means
that the wave function of two low-energy atoms is modified
by interactions even if the atoms are arbitrarily far apart.

The resonant limit is referred to as the unitary Fermi gas, because
the isotropic part of the scattering cross section is as large as
conservation of probability (unitarity) in quantum mechanics allows it
to be. The unitary Fermi gas is also scale invariant. This means
that physical observables are fixed by dimensional analysis and
universal functions of dimensionless ratios. We can apply this
type of argument to the sound diffusivity. On dimensional grounds,
$D$ is proportional to $\hbar/m$, where $\hbar$ is Planck's constant,
and $m$ is the mass of the atoms.

The constant of proportionality is determined by the detailed mechanism
for energy and momentum transfer. If momentum transfer is governed
by the diffusion of atoms then $D\sim \bar{p}l_{\it mfp}/m$, where
$\bar{p}$ is the mean momentum of an atom, and $l_{\it mfp}$ is the
mean free path. In a classical gas $\bar{p}l_{\it mfp}\gg \hbar$, but
we expect that in a strongly correlated gas the product of $\bar{p}$ and
$l_{\it mfp}$ is limited by quantum uncertainty, so that $D$ is
of order $\hbar/m$.

\begin{figure}[t]
\bc\includegraphics[width=8cm]{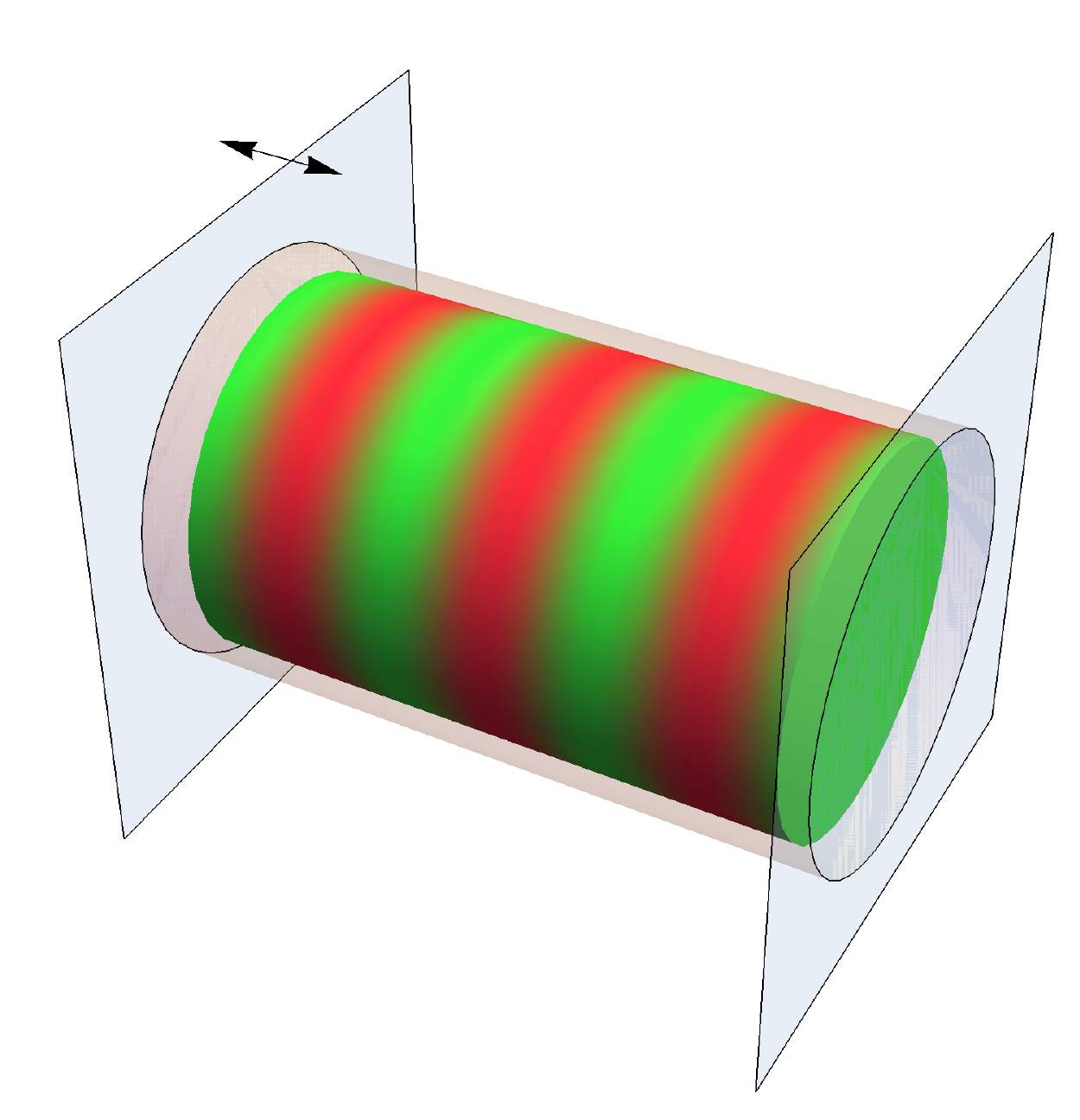}\ec
\caption{\label{fig_map}
Sound mode in a cylindrical box filled with a dilute atomic Fermi gas. 
A cylindrical box made of laser beams, 100 $\mu m$ long and 60 $\mu m$
in radius, contains about $2\cdot 10^6$ ultracold Lithium atoms. Sound
wave are excited by harmonically driving the intensity of the light
creating the endcaps. Sound diffusivity is measured by examining the
frequency width of the resonant standing waves. }   
\end{figure}

This is indeed what the results of Patel et al.~demonstrate. In the
experiment sound modes are excited by shaking the endcaps of the cylindrical
box. The position of resonances determined the speed of sound, and the width 
of the resonance determines the diffusivity. Patel et al.~find that the 
diffusivity drops as the temperature is lowered, and that it settles at a 
value around $D\simeq 1.5\hbar/m$ near the transition to a superfluid. This 
value is consistent with attempts to measure the shear viscosity and
thermal conductivity of the unitary Fermi gas individually
\cite{Bluhm:2017rnf,Baird:2019}, as well as with theoretical calculations
\cite{Enss:2010qh}. Below the critical temperature the unitary gas 
forms a superfluid which is roughly analogous to Bardeen-Cooper-Schrieffer 
(BCS) superconductivity, but with a parametrically large pairing gap and 
critical temperature. It is interesting to note that there are no sharp
features in the diffusivity at the phase transition temperature. 

The results of Patel et al.~have direct implications for the
structure of spinning neutron stars. The matter in the outer
layer of a neutron star (outside the core, but below the crust)
is a dilute liquid of neutrons. The neutron-neutron scattering
length is much larger than the distance between neutrons, so
that the results for the unitary Fermi gas are directly
applicable, even though the temperatures and densities are
many orders of magnitude larger. What matters is that dimensionless
ratios, such as the mean particle distance in units of the
thermal de Broglie wave length, are similar.

Neutron stars have many possible modes of oscillations. A
special class that arises due the Coriolis force in rotating stars
is known as Rossby, or simply r-modes. These modes are unstable,
and if they are not damped by momentum or energy diffusion then
large amplitude r-modes would lead to strong gravitational wave 
emission, and a rapid spin-down of the star. Understanding the 
diffusivity of neutron star matter is crucial to predicting the 
range of allowed spin frequencies, and possible r-mode signals
in gravitational wave detectors.

More generally the results of Patel et al.~shed light on the mechanism 
of transport in other strongly correlated quantum gases, such as the 
quark-gluon plasma investigated in heavy ion collisions at the 
Relativistic Heavy Ion Collider (RHIC) and the the Large Hadron Collider 
(LHC). The quark-gluon plasma is a state of matter that existed microseconds
after the Big Bang, at a temperature $T\simeq 2\cdot 10^{12}$ K. 
Measurements indicate that the momentum diffusivity of the quark 
gluon plasma is quite low. In a relativistic setting the mass of the 
particles is very small, and the natural scale for $D$ is $\hbar c^2/
(k_BT)$. Experiments based on the hydrodynamic expansion of the plasma
give values as small as $D\simeq 0.1 \hbar c^2/(k_BT)$. This number 
has been interpreted in terms of holographic models inspired by advances 
in string theory \cite{Kovtun:2004de}. However, in relativistic heavy ion
collisions the precise mechanism of momentum transport is difficult to
determine. This problem can potentially be tackled in future experiments
with cold gases, for example by carefully mapping the frequency dependence
of the response of the gas to external perturbations.


\end{document}